\documentclass{desyproc}
\usepackage{upgreek}

\begin{document}
\title{Using an InGrid Detector to Search for Solar Chameleons with CAST}

\author{{\slshape  Klaus Desch$^{1}$, Jochen Kaminski$^1$, Christoph Krieger$^1$, Michael Lupberger$^1$}\\[1ex]
$^1$University of Bonn, Bonn, Germany}

\contribID{familyname\_firstname}

\confID{11832}  
\desyproc{DESY-PROC-2015-02}
\acronym{Patras 2015} 
\doi  

\maketitle

\begin{abstract}
We report on the construction, operation experience, and preliminary background measurements
of an InGrid detector, i.e.~a MicroMegas detector with CMOS pixel readout. The detector was 
mounted in the focal plane of the Abrixas X-Ray telescope at the CAST experiment at CERN. 
The detector is sensitive to soft X-Rays in a broad
energy range (0.3--10~keV) and thus enables the search for solar chameleons. 
Smooth detector operation during CAST data taking in autumn 2014 has
been achieved. Preliminary analysis of background data indicates a background rate of 
$1-5\times 10^{-5}\,\mathrm{keV}^{-1}\mathrm{cm}^{-2}\mathrm{s}^{-1}$ above 2\,keV and 
$\sim 3\times 10^{-4}\,\mathrm{keV}^{-1}\mathrm{cm}^{-2}\mathrm{s}^{-1}$ around 1\,keV.
An expected limit of $\beta_\gamma \lesssim 5\times 10^{10}$ on the chameleon photon coupling
is estimated in case of absence of an excess in solar tracking data. We also discuss the prospects
for future operation of the detector.
\end{abstract}

\section{The CAST experiment}

The CERN Axion Solar Telescope (CAST)~\cite{Zioutas:1998cc} is operating since 2003 in search for the emission of
axions from the Sun through their conversion into soft X-Ray photons in the strong magnetic 
field of an LHC dipole prototype magnet. The experiment has been setting the strongest bounds on
solar axion production to date~\cite{Arik:2013nya}. More recently, CAST is extending its scope,
making use of the versatility of the experimental setup. These extensions include the search for
solar chameleons both through their coupling to photons~\cite{Anastassopoulos:2015yda} and through their coupling
to matter~\cite{CAST_KWISP} as well as the search for relic axions exploiting resonant microwave cavities 
immersed into the magnetic field~\cite{CAST_RELIC}. In these proceedings we report about the progress
in the search for solar chameleons using an InGrid detector, extending the preliminary results reported
at the 2014 Axion-WIMP workshop~\cite{Krieger:2014pea}.

\section{Solar Chameleons}

The observation of a non-vanishing cosmological constant, dubbed Dark Energy (DE), is arguably one of the
greatest mysteries of modern physics. There exist only very few particle physics approaches to explain DE. 
The observed accelerated expansion of the universe may be explained by the existence of a scalar field. 
One such scenario is the so-called chameleon for which a low-energy effective theory has been 
formulated~\cite{Chameleons}. The chameleon field acquires an effective mass through a screening potential
which establishes a non-zero vacuum expectation value depending on the surrounding matter density. 
The screening potential assures the suppression of measurable fifth force effects and leads to a
chameleon mass which depends on the ambient matter density. 
Chameleons, similar to axions, can be created via the Primakoff effect 
in strong electro-magnetic fields present in the Sun and observed on Earth through their back-conversion 
into detectable X-ray photons within a strong magnetic field via the inverse Primakoff effect. 
The energy of the photons is essentially equivalent to the chameleons' thermal energy during their production 
in the Sun. While axions may be created in the core of the Sun with a spectral maximum at approximately 3\,keV, 
chameleons are predicted to be created in the solar tachocline~\cite{Brax:2010xq} around 0.7\,R$_\odot$ 
where intense magnetic fields are present. Thus, they are produced at lower temperature 
corresponding to a spectral maximum of only 600\,eV, requiring photon detectors with sub-keV sensitivty. 
An initial search for solar chameleons with CAST has been conducted using a Silicon Drift 
Detector~\cite{Anastassopoulos:2015yda}. 

\section{InGrid Detector}

An InGrid (``Integrated Grid'') is a gas-ampficication device based on the MicroMegas principle. 
A thin aluminum mesh is mounted approximately 50 $\mu$m above a CMOS pixel chip, in our case the 
TimePix ASIC~\cite{timepix}, via photolithographic wafer post-processing techniques~\cite{ingrid-production}. 
The input pads of the pixels' charge-sensitive amplifiers serve as charge-collecting anodes and the collected
charged is amplified and processed digitally in-situ. The pixel pitch is 55$\times$55\,$\upmu$m$^{2}$.
With this fine pitch, a typical gas amplification of $\sim$3000 and a detection threshold of 
$\lesssim 1000$\,electrons, a single electron efficiency of $>$95\% is achieved. Given the diffusion of the
ionization electrons from the photo electron, this allows for the counting of the total number of 
created electrons on the pixel chip and yields a direct measure of the energy, free of fluctuations in
the amplification region. As the range of the photo electron in the detector gas (97.7\% Argon, 2.3\% Isobutane)
is only few hundred microns, the image of an absorbed photon is an essentially circular cloud of 
hit pixels, where the cloud radius decreases with the absorption depth of the photon. This pattern provides
an effective template which differs significantly from charged particle background (e.g.~cosmic muons or 
electrons from $\beta$-decay) which produces typically a track-like pattern on the pixel chip. 
These differences are exploited to provide a powerful topological background suppression. 
The detector and its installation in CAST is explained in more detail 
in~\cite{Krieger:2014pea, Krieger:2013cfa} where also sensitivity of the detector down to below 300\,eV
has been demonstrated.

\section{Results and Prospects}

\begin{figure}[htbp]
\centerline{\includegraphics[width=0.9\textwidth]{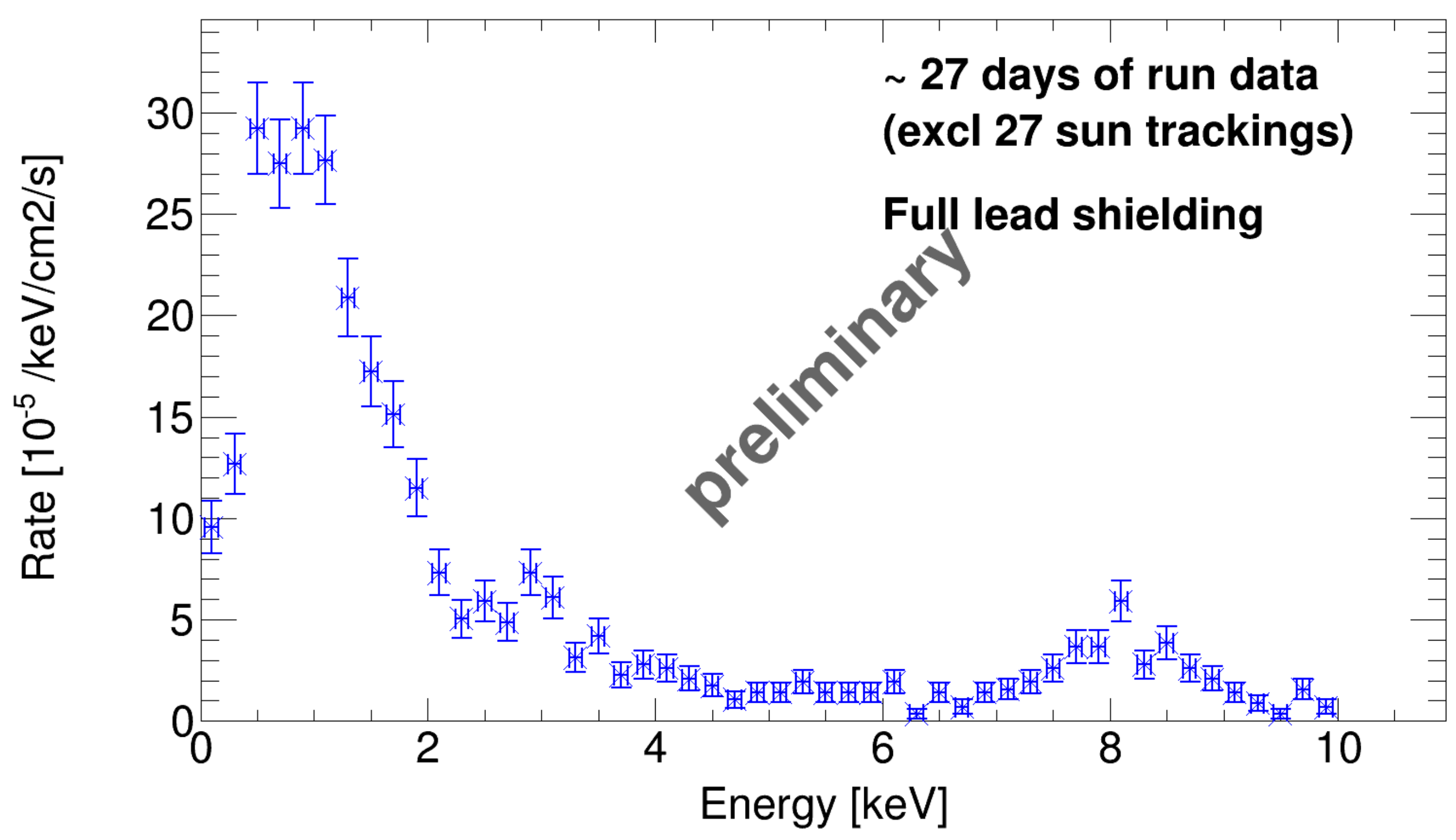}}
\caption{Preliminary background rate of InGrid Detector during operation in CAST in autumn 2014.}
\label{fig:background}
\end{figure}

In autumn 2014, the detector has, for the first time, been
taking data on 27~consective days including 1.5\,h of daily solar tracking. While the solar tracking data
are still blinded, the in-situ background data are being analysed using a simple three-variable
likelihood for the photon hypothesis. In comparison to~\cite{Krieger:2014pea}, the likelihood has
been further tuned to reduce energy-dependent biases. A preliminary background spectrum is shown in 
Fig.\,\ref{fig:background}. In the region above 2\,keV two peaks around 3\,keV and 8\,keV are visible.
The former corresponds to the known flourescence line of Argon while the latter is likely a 
superposition of Copper flourescence and cosmic tracks which traverse the detector parallel to the
drift field. Such cosmics produce a m.i.p. signal which, due to the track's direction, is difficult
to distinguish from a photon via topological supression alone. Outside these peaks, the background level is
around $1-2\times 10^{-5}\,\mathrm{keV}^{-1}\mathrm{cm}^{-2}\mathrm{s}^{-1}$. There is a notable 
increase in background for energies below 2\,keV, reaching 
$\sim 3\times 10^{-4}\,\mathrm{keV}^{-1}\mathrm{cm}^{-2}\mathrm{s}^{-1}$ around 1\,keV. The origin of
this background needs further study.

\begin{figure}[htbp]
\centerline{\includegraphics[width=0.8\textwidth]{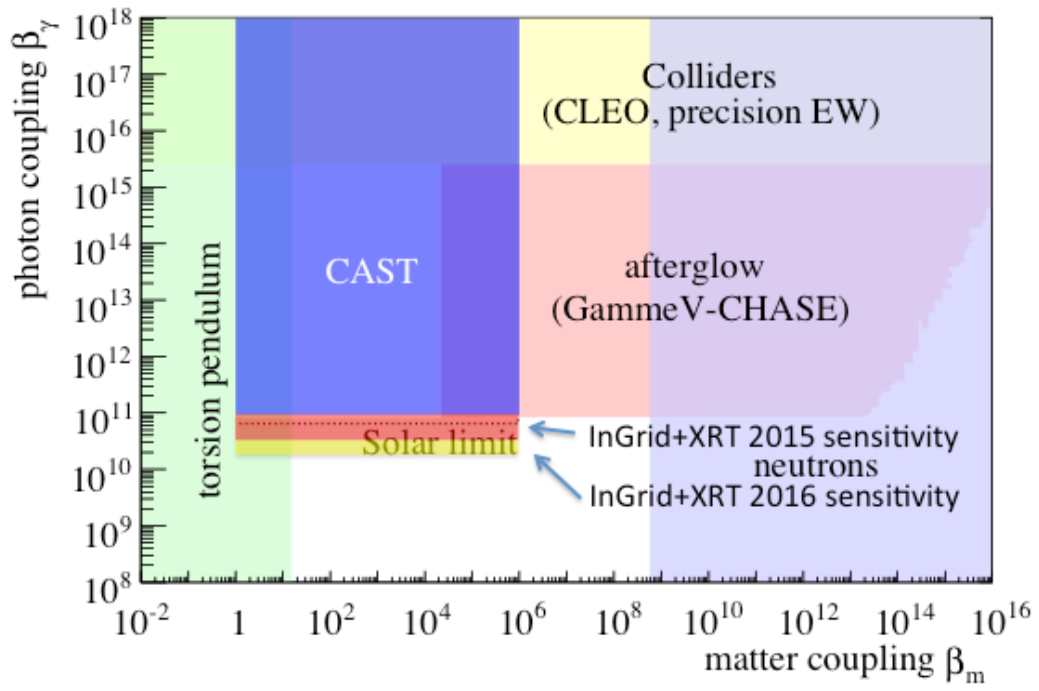}}
\caption{Exclusions in the plane of chameleon-matter coupling $\beta_m$ vs.~chameleon-photon coupling $\beta_\gamma$. Figure from~\cite{Anastassopoulos:2015yda} 
and modified to include InGrid detector prospects.}\label{fig:limit}
\end{figure}

While the solar tracking data of the 2014 run have not yet been analyzed, one can already
estimate an expected limit in case of non-observation of an excess. Our estimates are based on
scaling the limit of the SDD detector~\cite{Anastassopoulos:2015yda} and 
accounting for scaling factors in exposure time, effective sensitive area, background, and efficiency.
In Fig.\,\ref{fig:limit} the estimated expected limit of the chameleon-photon coupling $\beta_\gamma$ 
from the InGrid detector is shown together with the observed SDD limit and other experimental and 
astrophysical constraints. It can be seen that the 2014 InGrid data have the potential to set a limit
$\beta_\gamma \lesssim 5\times 10^{10}$, improving the SDD limit by almost a factor two under the same model assumptions as given in~\cite{Anastassopoulos:2015yda}. Also shown are prospects for data taking in 
2015 and 2016. At the time of writing, the detector has been continously taking data in the 2015 CAST run 
using the same setup as in 2014. Further improvements in background suppression (external cosmic veto,
additional readout of the grid signal) and photon detector efficiency (through thinner X ray windows)
as well as improvements in the software rejection of background are currently being developed and will
be implemented step-wise. Rough estimates for the potential of these improvements yield expected exclusions
are also shown in Fig.\,\ref{fig:limit}.

\section{Acknowledgments}

We thank the organizers of the Axion-WIMP-Workshop 2015 for an exciting conference and their warm
hospitality.

\section{Bibliography}


\begin{footnotesize}

\end{footnotesize}


\end{document}